\documentclass{article}

\usepackage[english]{babel}

\usepackage[letterpaper,top=2cm,bottom=2cm,left=3cm,right=3cm,marginparwidth=1.75cm]{geometry}

\usepackage{amsmath, amssymb}
\usepackage{graphicx}
\usepackage[colorlinks=true, allcolors=blue]{hyperref}
\usepackage[
    style=authoryear, 
    sorting=nty, 
    maxbibnames=99, 
    minbibnames=99,
    maxcitenames=2,
    mincitenames=1,
    uniquelist=false
]{biblatex}
\usepackage{algorithm}
\usepackage{algpseudocode}

\title{On skew-symmetric distributions and their use in Monte Carlo sampling algorithms: coordinate-free, Gibbs-style and manifold versions of the Barker proposal}
\author{
Minh Vu\thanks{Corresponding author. E-mail: \href{mailto:vungocminh.nmv@gmail.com}{vungocminh.nmv@gmail.com}}\;$^{1}$, 
Samuel Livingstone$^{2}$ \& 
Pantelis Samartsidis$^{3}$ \\
\\
$^{1}$Department of Statistics, University of Oxford \\
$^{2}$Department of Statistical Science, University College London \\
$^{3}$MRC Biostatistics Unit, University of Cambridge
}

\begin{document}
\maketitle

\begin{abstract}
Skew-symmetric probability distributions provide a principled mechanism for incorporating gradient information into Markov chain Monte Carlo algorithms. Here we review the (preconditioned) Barker proposal, a Metropolis–Hastings algorithm built on skew-symmetric distributions, and motivate its design. We then introduce three natural extensions. First, we propose coordinate-free variants of the Barker algorithm. Second, we introduce a Gibbs-style Barker algorithm that re-evaluates the gradient at each partially updated coordinate. Third, we derive a simplified manifold Barker algorithm, producing a manifold sampler with enhanced robustness compared to natural comparators. Numerical experiments demonstrate that the Gibbs-style variant improves raw sampling efficiency on correlated targets, that the coordinate-free variants offer limited practical advantage over the standard Barker proposal once computational costs are accounted for, and that the simplified manifold Barker algorithm can achieve significant advantages over simplified manifold MALA when the local geometric structure of the target is irregular or unreliable.
\end{abstract}


\section{Introduction}

Monte Carlo sampling algorithms are a ubiquitous tool in modern probabilistic modeling. The need to compute expectations that are not available analytically is commonplace in the classical settings of Bayesian inference and maximum intractable likelihood estimation, as well as the more contemporary settings of deep generative models and reinforcement learning.

Sampling algorithms based on Markov chains are the gold standard for high-dimensional Monte Carlo.  The Metropolis--Hastings algorithm provides an umbrella approach from which to design more specific methods, and reduces the question of designing a useful algorithm to that of designing an effective proposal at each iteration of the algorithm, which is then accepted or rejected using the Metropolis--Hastings filter.  When sampling from continuous probability distributions with smooth densities, state of the art approaches use the gradient of the log-density to guide proposals into favorable regions \parencite{green2015bayesian, vstrumbelj2024past}. This strategy has proven highly successful in practice and is well motivated by the study of continuous-time Markov processes and their mixing properties \parencite{green2015bayesian}.

Skew-symmetric probability distributions were only recently suggested as a choice of proposal in a Metropolis--Hastings algorithm, via the Barker proposal of \textcite{livingstone2022barker}.  The skewing mechanism provides a principled way to introduce the gradient, leading to an approach that can work well when other gradient-based methods fail.  The skew-symmetric proposal can be theoretically motivated in two different ways, either as an approximation to the transition density of a stochastic differential equation \parencite{iguchi2026skew}, or as approximating the dynamics of a Markov jump process \parencite{livingstone2025foundations,hird2020fresh}.  In both cases the process being approximated has limiting distribution equal to that from which samples are desired.

In this article we will review skew-symmetric distributions in Markov chain Monte Carlo sampling and motivate their use both theoretically and practically.  We then introduce some natural extensions and evaluate them on some benchmark examples.  In Section \ref{sec:barker_proposal} we review the Barker proposal scheme as a Metropolis--Hastings algorithm, paying close attention to the role of skew-symmetric distributions in its design.  In Sections \ref{subsec:motivation_langevin}-\ref{subsec:motivation_lb} we discuss the theoretical motivation behind the algorithm, via its connection to two continuous-time stochastic processes, the overdamped Langevin diffusion and the locally-balanced Markov jump process.  

In Section \ref{sec:barker_coordinatefree} we introduce a new version of the Barker proposal with more natural geometric properties, defined as a mixture of (element-wise) skew-symmetric kernels over orthogonal transformations. We consider a version obtained by averaging with respect to the Haar measure over the orthogonal group, and another using a more computationally efficient discrete mixture distribution based on Walsh--Hadamard transforms \parencite{ailon2009fast,tropp2011improved}.  In Section \ref{sec:barker_gibbs} we introduce another perturbation of the Barker proposal in which the skew-symmetric kernels are defined using iterative computation of the gradient vector.  This algorithm, termed `Gibbs-style Barker', can exploit intermediate decisions to guide the proposal in a more informed way than the standard Barker algorithm when the distribution of interest has strong dependencies between coordinates.  We compare each of these new algorithms to the standard Barker proposal on some natural benchmark models.

In Section \ref{sec:barker_manifold} we consider skew-symmetric distributions arising from approximation of more general diffusion process dynamics, which naturally leads to an algorithm that exploits not only gradients but second and third derivative information about the distribution of interest when applied to the overdamped Langevin diffusion with position-dependent volatility \parencite{xifara2014langevin}.  The class of algorithms based on this object and other related stochastic processes is called \emph{manifold methods}.  We introduce a simplified manifold Barker algorithm, following the approach of the simplified manifold MALA algorithm of \textcite{girolami2011riemann}, in which terms involving third derivatives are ignored to reduce computational costs.  We then compare simplified manifold Barker and MALA on some numerical examples to illustrate the enhanced robustness properties that the skew-symmetric approach provides in the setting of manifold Markov chain Monte Carlo.







\section{The Barker proposal}
\label{sec:barker_proposal}

Suppose we are given the task of computing expectations with respect to a distribution on $\mathbb{R}^d$ with density $\pi$ (which need not be normalized).  Given a current state $X^{(n)} \in \mathbb{R}^d$ for some $n \in \mathbb{N}$, a single iteration of the Metropolis--Hastings algorithm proceeds in two steps.  First, we generate a proposed next state $Y|X^{(n)}$ from  a conditional distribution with density $q(\cdot|X^{(n)})$.  We then set $X^{(n+1)} = Y$ with probability 
$$
\alpha(X^{(n)},Y) = \min\left(1, \frac{\pi(Y)q(X^{(n)}|Y)}{\pi(X^{(n)})q(Y|X^{(n)})}\right),
$$
otherwise setting $X^{(n+1)} = X^{(n)}$.  Beginning from an initial state $X^{(0)}$, these steps can be iterated to produce a Markov chain $\{X^{(0)},X^{(1)},X^{(2)},...\}$.  It is a standard exercise to show that this Markov chain will satisfy the detailed balance equations with respect to $\pi$, meaning $\pi$ will be the limiting distribution under mild regularity conditions and ergodic averages of the form $m^{-1}\sum_{n=0}^{m-1} h(X^{(n)})$ can be used to approximate the integral $\mathbb{E}_\pi[h(X)]$ for some appropriate function $h$.

A simple choice for $q$ is a random walk, meaning that $q(x+z|x) = f_d(z)$, where $f_d$ is a centered, symmetric probability density on $\mathbb{R}^d$.  The resulting algorithm is called the random walk Metropolis, and is both celebrated and well-studied \parencite{sherlock2010random,green2015bayesian}.  It is known, however, that more sophisticated choices of $q$ can lead to faster mixing Markov chains when the dimension $d$ is large.  One tool that is often used in the design of $q$ is $\nabla\log\pi(x)$, the gradient of $\log\pi$, which is available in closed form even when only $c\pi$ is known for some unknown $c >0$.  This is often the case in applications such as Bayesian inference.  This gradient term appears in the dynamics of several $\pi$-invariant continuous-time Markov processes, which can be used as inspiration for the design of $q$.  More details of such processes are given in Section \ref{subsec:motivation_langevin}.  

The (preconditioned) Barker proposal is the choice of proposal density
\begin{equation} 
\label{eq:barker_proposal}
q(y|x) \propto \prod_{i=1}^d 2 \cdot F \left(c_F \cdot [L^\top\nabla\log\pi(x)]_i \cdot z_i\right) \cdot f_\sigma(z_i), \quad z := L^{-1}(y - x),
\end{equation}
where $f_\sigma$ is a symmetric probability density function on $\mathbb{R}$ with scale $\sigma > 0$, $F$ is a (differentiable) cumulative distribution function of a symmetric random variable on $\mathbb{R}$, and $c_F := 1/(4F'(0))$. Each component of the proposal $Y$ is therefore a skew-symmetric random variable, with location parameter $x$, scale determined by the preconditioner $L$, and skewness parameter $c_F [L^\top\nabla\log\pi(x)]_i$, proportional to the $i$th component of the preconditioned gradient $L^\top\nabla\log\pi(x)$ at $x$. The original algorithm of \textcite{livingstone2022barker} sets $f_\sigma$ to be a Gaussian density and $F$ to be the logistic function, for reasons explained in the last paragraph of this subsection. Another appeal of the logistic choice of $F$ is that $c_F = 1$.  It was later found that choosing $f_\sigma$ to be a bimodal distribution can increase the efficiency of the algorithm \parencite{vogrinc2023optimal}. Another natural choice of $F$ is the standard Gaussian CDF, meaning \eqref{eq:barker_proposal} becomes a unified skew-Normal density \parencite{arellano2006unification}.

Simulating from \eqref{eq:barker_proposal} is straightforward \parencite{azzalini2013skew}.  A detailed recipe for a single iteration is provided as Algorithm \ref{alg:mhbarker}.  

\begin{algorithm}
\caption{Preconditioned Barker}
\label{alg:mhbarker}
\begin{algorithmic}[1]
\Require Current state $X^{(n)} = x \in \mathbb{R}^d$, step size $\sigma$, preconditioner $L \in \mathbb{R}^{d \times d}$
\For{$i \in \{1, \ldots, d\}$} draw $z_i \sim f_\sigma$ 
\EndFor
\For{$i \in \{1, \ldots, d\}$} draw $b_i \sim \mathrm{Rademacher}\!\left(F\!\left(c_F \cdot [L^\top \nabla\log\pi(x)]_i \cdot z_i\right)\right)$
\EndFor
\State Set $y = x + L(b \odot z)$ (where $\odot$ denotes the element-wise product)
\State Set $X^{(n+1)} = y$ with probability $\alpha(x, y)$, otherwise set $X^{(n+1)} = x$, where
\begin{equation*}
    \alpha(x, y) = \min\!\left(1,\; \frac{\pi(y)}{\pi(x)} \cdot \prod_{i=1}^d \frac{1 - F\!\left(c_F \cdot b_i z_i [L^\top \nabla\log\pi(y)]_i\right)}{F\!\left(c_F \cdot b_i z_i [L^\top \nabla\log\pi(x)]_i\right)}\right)
\end{equation*}
\end{algorithmic}
\end{algorithm}

The algorithm inherits an important stability property from skew-symmetric distributions.  The size of the proposal $|y_i - x_i| = |z_i|$, where $z_i \sim f_\sigma$ is a draw from the symmetric base density $f_\sigma$.  Crucially, therefore, the size of the increment is not influenced by the magnitude of the gradient vector.  This is in stark contrast to alternative gradient-based approaches, in which the size of the proposed move is essentially proportional to the size of the gradient when the latter is large.  Because of this, gradient-based algorithms typically become unstable when the gradient is not a Lipschitz function of $x$ or behaves irregularly in some other way.  Alternative approaches to combating this phenomenon tend to modify the algorithm so that only the unit vector in the gradient direction is used, but this sacrifices the information about the size of the gradient, which is informative if used correctly.  The skew-symmetric construction of the Barker algorithm, by contrast, uses both the magnitude and the direction of the gradient, but does so in a way that does not result in stability issues when the gradient is very large or otherwise irregular.  In practical applications, it seems that this robustness is most easily seen when the global scale $\sigma$ and the preconditioner $L$ are learned adaptively during the sampling (e.g. \textcite{andrieu2008tutorial}).  The optimization schemes used to choose $\sigma$ and $L$ seem to converge much more quickly in the presence of irregular gradients for the Barker proposal as compared to other gradient-based sampling algorithms, and this can be justified theoretically by studying spectral gaps of the resulting Markov chains \parencite{livingstone2022barker}.

The name `Barker' proposal pays homage to an alternative Metropolis--Hastings style algorithm proposed by the physicist AA Barker \parencite{barker1965monte}.  In place of the acceptance rate $\min(1,t)$, where t is the Hastings ratio $\pi(y)q(x|y)/(\pi(x)q(y|x))$, the function $1/(1+t^{-1})$, sometimes called `Barker's rule', is used.  When $q(y|x) = q(x|y)$ (as in the random walk Metropolis), this leads to an algorithm in which the acceptance rate for a proposal becomes
\[
\alpha_B(x,y) = \frac{1}{1 + \pi(x)/\pi(y)}.
\]
If $x \in \mathbb{R}$ then a first-order approximation gives
\begin{equation}
\label{eq:barker_connection}
\frac{\pi(x)}{\pi(y)} \approx \exp \left( -(y-x)\cdot \frac{d}{dx}\log\pi(x) \right) \implies \alpha_B(x,y) \approx \frac{1}{1+\exp\left( -(y-x)\cdot \frac{d}{dx}\log\pi(x)\right)},
\end{equation}
which is precisely the logistic cumulative distribution function evaluated at the same value as in \eqref{eq:barker_proposal} when $L =I$.  The algorithm can therefore be viewed as acting in two stages.  First an initial proposal $Y^* = x+Z$ is generated, where $Z \sim f_\sigma$ and $X^{(n)} = x$. Then an accept-reject decision is made based on the approximation \eqref{eq:barker_connection}. If this first stage results in acceptance then we set the final proposal $Y= Y^*$ and accept or reject it as described in Algorithm \ref{alg:mhbarker}.  If this first stage results in rejection, however, the alternative final proposal $Y = x - Z$ is then used in place of $Y^*$.  This is connected to delayed-acceptance Metropolis--Hastings approaches, in which a proposal must also pass through two stages of accept-reject decisions to be ultimately accepted, and in which the first stage acceptance rate is often some approximation to the usual Metropolis--Hastings rule (see e.g. \textcite{banterle2015accelerating}).

\subsection{Motivation from Langevin dynamics}
\label{subsec:motivation_langevin}

One example of a continuous-time process that has limiting distribution $\pi$ under mild conditions is the (preconditioned) overdamped Langevin diffusion, the solution to the stochastic differential equation
\begin{equation}
\label{eq:langevin_diffusion}
dX(t) = \frac{1}{2}A\nabla\log\pi(X(t))dt + \sqrt{A}\,dW(t), \quad X(0)  = x,
\end{equation}
for some positive-definite matrix $A \in \mathbb{R}^{d \times d}$.  This process is the inspiration for numerous sampling algorithms, both within the Metropolis--Hastings class and beyond (e.g. \textcite{durmus2019high,welling2011bayesian}).  A natural way to design a Metropolis--Hastings proposal based on \eqref{eq:langevin_diffusion} is to approximate the intractable transition density $p_s(\cdot|x)$ meaning the density of $X(t+s)|(X(t) = x)$, with some tractable surrogate $\tilde{p}_s(\cdot|x)$ over a small time period $s >0$.  Repeatedly simulating $X^{(n+1)} \sim \tilde{p}_s(\cdot|X^{(n)})$ will produce a valid Markov chain without further modification, but it will not in general have limiting distribution $\pi$ because $\tilde{p}_s$ is not exactly the same as $p_s$. To correct for this, we can instead set $\tilde{p}_s(\cdot|x) = q(\cdot|x)$, the proposal in a Metropolis--Hastings algorithm, and allow the Metropolis--Hastings filter $\alpha$ to ensure that the Metropolized Markov chain will have limiting distribution $\pi$.

The most conventional way to design Metropolis--Hastings proposals using \eqref{eq:langevin_diffusion} is via the Euler--Maruyama discretization scheme \parencite{kloeden1989survey}.  This equates to approximating the transition density $p_s(\cdot|x)$ with a Gaussian of the form $N(x + sA\nabla\log\pi(x)/2, sA)$.  Choosing the proposal density $q$ in this way produces the Metropolis-adjusted Langevin algorithm, or MALA \parencite{roberts1996exponential}.

An alternative numerical scheme for stochastic differential equations was recently proposed in \textcite{iguchi2026skew}, in which the approximation for $p_s$ is built from a skew-symmetric distribution.  For a general stochastic differential equation of the form $dX(t) = \mu(X(t))dt + \Sigma(X(t))^{1/2} dW(t)$ on $\mathbb{R}^d$ with $\Sigma :\mathbb{R}^d \to \mathbb{R}^{d \times d}$, the skew-symmetric numerical scheme consists of generating a skeleton path $\{Y^{(0)}, Y^{(1)},...,\}$ of the process using the recursion $Y^{(n+1)}|Y^{(n)} \sim \tilde{p}_s(\cdot|Y^{(n)})$, where
\begin{equation}
\label{eq:skewsymmetric_scheme}
\tilde{p}_s(y|x) \propto \prod_{i=1}^d 2\cdot F\left( 2c_F \cdot [\Sigma^{-1/2}(x)\mu(x)]_i \cdot z_i \right)\cdot f_{s^{1/2}}\left(z_i \right), \quad z := \Sigma^{-1/2}(x)(y - x)
\end{equation}
with $F$ again the cumulative distribution of a symmetric random variable, $f_{s^{1/2}}$ a symmetric density with scale $s^{1/2}$ and as in the previous section $c_F := 1/(4F'(0))$ where $F'$ is the density associated with $F$.  The numerical scheme therefore consists of generating skew-symmetric increments, with the level of skew in each component of the state vector determined by the quantity $[\Sigma^{-1/2}(x)\mu(x)]_i$ multiplied by the factor $c_F$.  Intuitively, the skewness plays the role of biasing the increments in the direction of the vector $\Sigma^{-1/2}(x)\mu(x)$.  It is shown in \textcite{iguchi2026skew} that this choice leads to a valid numerical approximation to the solution of \eqref{eq:langevin_diffusion}, in the sense that as $s \to 0$ the true solution is recovered in the weak sense and that the scheme has weak order 1 under mild assumptions on the drift and volatility (e.g. \textcite{milstein2004stochastic}).  Further work in \textcite{iguchi2026pathwise} shows that the scheme also converges strongly in mean square when both $F$ and $f_{s^{1/2}}$ are Gaussian.

Substituting the coefficients of \eqref{eq:langevin_diffusion} into \eqref{eq:skewsymmetric_scheme} gives
\[
2c_F\cdot [\Sigma^{-1/2}(x)\mu(x)]_i = c_F\cdot [A^{1/2}\nabla\log\pi(x)]_i.
\]
Writing $A^{1/2} = L$, then setting $\sqrt{s} = \sigma$ then shows the equivalence of \eqref{eq:skewsymmetric_scheme} and \eqref{eq:barker_proposal}.  Choosing $F(x) = (1+e^{-x})^{-1}$, the logistic function, gives $F'(0) = 1/4$, meaning $c_F = 1$ and the skewness factor reduces to simply $[A^{1/2}\nabla\log\pi(x)]_i$.  



\subsection{Motivation from locally-balanced Markov processes}
\label{subsec:motivation_lb}

Another way to understand the Barker proposal is as an approximation to the transition density of a \emph{locally-balanced Markov jump process} \parencite{livingstone2025foundations,power2019accelerated}.  This is defined as a continuous-time pure jump type Markov process in which at current state $X(t) = x$, the next jump location is generated from a distribution with density
\[
j(y|x) = \frac{1}{\lambda(x)} \cdot g\left( \frac{\pi(y)q(x|y)}{\pi(x)q(y|x)}\right)q(y|x),
\]
where
\[
\lambda(x) := \int g\left( \frac{\pi(y)q(x|y)}{\pi(x)q(y|x)}\right)q(y|x)dy
\]
is chosen so that $j(\cdot|x)$ is appropriately normalized. Here $\pi$ and $q$ are densities as in the Metropolis--Hastings algorithm, and the function $g : \mathbb{R} \to \mathbb{R}$ must satisfy the functional equation $g(\tau) = \tau g(1/\tau)$ for all $\tau > 0$.  The next jump time is generated according to an exponential distribution with rate parameter $\lambda(x)$.

This process has limiting distribution $\pi$ under mild conditions \parencite{livingstone2025foundations}, and is in many ways similar to a continuous-time version of the Metropolis--Hastings algorithm, but also has some differences, notably that the function $g$ can be unbounded above, and that the jump rate is state-dependent.  Nonetheless, setting $g(\tau) := 1/(1+\tau^{-1})$ as before and making the same first-order approximation as in \eqref{eq:barker_connection} leads to the approximate jump density being precisely the Barker proposal \eqref{eq:barker_proposal} and the normalizing constant reducing to simply $1/2$, meaning the jumps no longer occur at a state-dependent rate in this first-order approximation to the locally-balanced process.  More details are given in Section 3 of \textcite{hird2020fresh}.

\section{A coordinate-free Barker algorithm}
\label{sec:barker_coordinatefree}

One criticism of the Barker proposal is that its construction depends on the choice of coordinate system $\{e_1,...,e_d\}$, via the partial derivatives $\partial_i\log\pi(x) = \nabla\log\pi(x)^\top e_i$ used to construct the skewness in the proposal.  We describe how this can be problematic in the next section, before giving a formal definition of a coordinate-free algorithm and then introducing a new coordinate-free Barker proposal in Section \ref{subsec:barker_randomized}.

\subsection{Problems caused by over-reliance on the coordinate system}
Sampling algorithms that are highly dependent on the choice of coordinate basis can be undesirable.  A simple example is a two-dimensional Gibbs sampler for the distribution $N(0, R_\theta DR_\theta ^\top)$, where
\[
R_\theta = \begin{pmatrix} \cos(\theta) & -\sin(\theta) \\ \sin(\theta) & \cos(\theta) \end{pmatrix}, \quad D = \begin{pmatrix} \sigma_1^2 & 0 \\ 0 & \sigma_2^2 \end{pmatrix},
\]
with $\theta \in (0,2\pi)$ and $\sigma_1 \neq \sigma_2$.  The systematic scan Gibbs sampler defined on this basis performs iterative sampling from the conditional distributions
\begin{align*}
X_1|(X_2=x_2) 
&\sim 
N\left(\frac{\cos(\theta)\sin(\theta)(\sigma_1^2-\sigma_2^2)}{\sin^2(\theta)\sigma_1^2 + \cos^2(\theta)\sigma_2^2}x_2 , \frac{\sigma_1^2\sigma_2^2}{\sin^2(\theta)\sigma_1^2 + \cos^2(\theta)\sigma_2^2}\right), 
\\
X_2|(X_1 = x_1) 
&\sim 
N\left(\frac{\cos(\theta)\sin(\theta)(\sigma_1^2-\sigma_2^2)}{\cos^2(\theta)\sigma_1^2 + \sin^2(\theta)\sigma_2^2}x_1 , \frac{\sigma_1^2\sigma_2^2}{\cos^2(\theta)\sigma_1^2 + \sin^2(\theta)\sigma_2^2}\right).
\end{align*}
The relaxation time for the algorithm is proportional to $1/(1-\rho^2)$, where
\[
\rho = \frac{\sin(\theta)\cos(\theta)(\sigma_1^2 -\sigma_2^2)}{\sqrt{(\sin^2(\theta)\sigma_1^2 + \cos^2(\theta)\sigma_2^2)(\cos^2(\theta)\sigma_1^2 + \sin^2(\theta)\sigma_2^2)}}.
\]
This relaxation time can be made arbitrarily large by choosing $\theta$, $\sigma_1$ and $\sigma_2$ appropriately, resulting in a sampler that converges to equilibrium very slowly \parencite{roberts1997updating}. If instead the change of variables $x' = R_{-\theta}x$ is applied, then the new distribution of interest is $N(0, D)$, for which the Gibbs sampler
\[
X_1'|(X_2'=x_2') \sim N(0, \sigma_1^2), \quad X_2'|(X_1' = x_1') \sim N(0, \sigma_2^2)
\]
consists of independent sampling from the two marginal distributions for $X_1'$ and $X_2'$, and reaches equilibrium in a single iteration.

The Barker proposal does not suffer from these kinds of issues in the same manner as the Gibbs sampler in the experience of the authors.  Nonetheless, performance can be affected by the coordinate system, and so it is desirable to see if this issue can be addressed.

\subsection{Coordinate-free Metropolis--Hastings proposals}

We will call a Metropolis--Hastings proposal coordinate-invariant or coordinate-free if its proposal transition density $q_{\pi,L}$ (which may depend on both $\pi$ and a preconditioner $L \in \mathbb{R}^{d \times d}$) satisfies the relationship 
\begin{equation}
\label{eq:coordinate_invariance}
q_{\pi_Q,QLQ^\top}(Qy|Qx) = q_{\pi,L}(y|x), \quad \pi_Q(x) := \pi(Q^\top x)
\end{equation}
for all $x$ and $y$ in $\mathbb{R}^d$, and any orthogonal matrix $Q \in \mathbb{R}^{d\times d}$.  Intuitively this means that changing the coordinate system will not affect where the proposal places its probability mass.  The Barker proposal does not satisfy \eqref{eq:coordinate_invariance}. A simple attempt to define a coordinate-free alternative is the skew-symmetric distribution
\[
q(y|x) \propto 2 \cdot F(c_F\cdot L^\top \nabla\log\pi(x)^\top z)f_\sigma(z), \quad z := L^{-1}(y-x).
\]
This is arguably a more geometrically natural choice of multivariate proposal, but was already considered in \textcite{livingstone2022barker} and shown to be inferior to \eqref{eq:barker_proposal}. Intuitively the limitation is that the information provided by the gradient is summarized in a single scalar $L^\top \nabla\log\pi(x)^\top z$.  This is discussed in some detail in Section 4 of \textcite{hird2020fresh}.  Another option is to consider a coordinate system based on the gradient itself, meaning that $\nabla \log\pi(x)$ is chosen as the first basis vector at each iteration.  In fact, however, this results in exactly the same coordinate-invariant proposal as that described above, though some algebra is required to see this.

\subsection{A randomized basis Barker proposal}
\label{subsec:barker_randomized}

Another way to achieve coordinate invariance is via randomizing the coordinate basis at each iteration.  Mathematically this results in a proposal density that is defined as a mixture of skew-symmetric distributions, given by
\begin{equation}
\label{eq:mixture_barker}
q_\mu(y|x) \propto \mathbb{E}_\mu\left[ \prod_{i=1}^d 2\cdot F(c_F \cdot [L^\top Q^\top \nabla\log\pi(x)]_iz_i)\cdot f_\sigma(z_i) \right], \quad z := L^{-1}Q^\top (y-x)
\end{equation}
where $\mu$ is a distribution defined over orthogonal matrices $Q \in \mathbb{R}^{d\times d}$.
This achieves coordinate invariance via randomizing the basis at each iteration.  Unfortunately, such a distribution will rarely be tractable. We can, however, generate a sample $Q \sim \mu$, and then generate a Metropolis--Hastings proposal using the conditional proposal density
\[
q_Q(y|x) \propto \prod_{i=1}^d 2\cdot F(c_F \cdot [L^\top Q^\top \nabla\log\pi(x)]_iz_i)\cdot f_\sigma(z_i), \quad z := L^{-1}Q^\top (y-x).
\]
The difference between this second approach and using \eqref{eq:mixture_barker} directly is that in the second approach the accept-reject decision is made based on $q_Q$, whereas in the first it is made based on $q_\mu$.  These can be directly related as the augmented and marginalized kernels with respect to the parameter $Q$, as discussed in \textcite{titsias2018auxiliary,grazzi2026randomized}.  It is shown in \textcite{titsias2018auxiliary} that using the marginalized kernel is always advantageous asymptotically in terms of reducing the variance of ergodic averages, but numerical results in \textcite{grazzi2026randomized} suggest that often the two approaches perform similarly in practice.  Crucially, the augmented approach using $q_Q$ is straightforward to implement provided that samples can be generated from $\mu$, and the additional cost of the scheme compared to the standard Barker proposal is dictated by how expensive this operation is combined with the additional multiplication by $Q^\top $ required to generate a proposal using $q_Q$.

\subsection{Two different strategies for sampling orthogonal matrices}
\label{subsec:orthogonal_matrix_sampling}


We consider two strategies for generating orthogonal matrices $Q \sim \mu$, which differ both in their computational cost and in the choice of $\mu$. 

The first strategy takes $\mu$ to be the uniform distribution over the space of $d \times d$ orthogonal matrices, defined via the Haar measure. In the absence of any informed preference for a particular basis, this is a natural default. Samples are drawn using a QR-based algorithm \parencite{mezzadri2007generate}, at a cost of $O(d^3)$ per draw and $O(d^2)$ per subsequent matrix-vector product.

The second strategy generates a structured random orthogonal matrix $Q = HD$, where $H$ is the normalized Walsh--Hadamard matrix defined recursively by
\begin{equation*}
    H_{2n} = \frac{1}{\sqrt{2}} \begin{pmatrix} H_n & H_n \\ H_n & -H_n \end{pmatrix}, \qquad H_1 = 1,
\end{equation*}
and $D$ is a diagonal matrix of i.i.d. Rademacher signs \parencite{ailon2009fast}. Since $H^\top H = I$ and $D^\top D = I$, $Q = HD$ is orthogonal by construction. When $d$ is not a power of two, the target is zero-padded to the next power of two before applying the transform. The distribution of $Q$ is therefore uniform over the discrete set of $2^d$ orthogonal matrices obtained by independently sign-flipping each column of $H$. By contrast, the Haar measure is continuous and assigns positive probability to every open subset of the orthogonal group. 

In our implementation, $H$ is constructed once as a dense matrix and all randomness enters through $D$. Sampling $D$ costs $O(d)$, and applying $Q$ by standard matrix-vector multiplication costs $O(d^2)$. The Barker samplers using each strategy are given in Algorithm~\ref{alg:rotated_barker} and Algorithm~\ref{alg:hadamard_barker}, and their performance is compared empirically in Section \ref{sec:numerics}. Here and in the next section we set the preconditioner $L = I$ for simplicity.

\begin{algorithm}
\caption{Rotated Barker}
\label{alg:rotated_barker}
\begin{algorithmic}[1]
\Require Current state $X_n = x \in \mathbb{R}^d$, step size $\sigma$
\State Draw $Q \sim \mathrm{Haar}(d)$
\For{$i \in \{1, \ldots, d\}$} draw $z_i \sim f_\sigma$
\EndFor
\For{$i \in \{1, \ldots, d\}$} draw $b_i \sim \mathrm{Rademacher}\!\left(F\!\left(c_F \cdot [Q^\top \nabla\log\pi(x)]_i \cdot z_i\right)\right)$
\EndFor
\State Set $y = x + Q(b \odot z)$
\State Set $X_{n+1} = y$ with probability $\alpha(x, y)$, otherwise set $X_{n+1} = x$, where
\begin{equation*}
    \alpha(x, y) = \min\!\left(1,\; \frac{\pi(y)}{\pi(x)} \cdot \prod_{i=1}^{d} \frac{1 - F\!\left(c_F \cdot b_i z_i [Q^\top \nabla\log\pi(y)]_i\right)}{F\!\left(c_F \cdot b_i z_i [Q^\top \nabla\log\pi(x)]_i\right)}\right)
\end{equation*}
\end{algorithmic}
\end{algorithm}

\begin{algorithm}
\caption{Hadamard Barker}
\label{alg:hadamard_barker}
\begin{algorithmic}[1]
\Require Current state $X_n = x \in \mathbb{R}^d$, step size $\sigma$, padded dimension $\tilde d = 2^{\lceil\log_2d\rceil}$, normalized Hadamard matrix $H_{\tilde{d}}$
\For{$i \in \{1, \ldots, \tilde{d}\}$} draw $D_i \sim \text{Rademacher}(1/2)$ 
\EndFor
\For{$i \in \{1, \ldots, \tilde{d}\}$} draw $z_i \sim f_\sigma$
\EndFor
\For{$i \in \{1, \ldots, \tilde{d}\}$} draw $b_i \sim \text{Rademacher}\left( F \left(c_F \cdot \left[ D \odot H^\top \begin{pmatrix} \nabla \log \pi(x) \\ \mathbf{0}_{\tilde{d}-d} \end{pmatrix} \right]_i \cdot z_i \right) \right)$
\EndFor
\State Set $y = x + \left[H \left( D \odot b \odot z \right) \right]_{1:d}$
\State Set $X_{n+1} = y$ with probability $\alpha(x, y)$, otherwise set $X_{n+1} = x$, where
\begin{equation*}
    \alpha(x, y) = \min\!\left(1,\; \frac{\pi(y)}{\pi(x)} \cdot \prod_{i=1}^{\tilde{d}} \frac{1 - F\!\left(c_F \cdot b_i z_i \left[D \odot H^\top\begin{pmatrix}\nabla\log\pi(y) \\ \mathbf{0}_{\tilde{d}-d}\end{pmatrix}\right]_i\right)}{F\!\left(c_F \cdot b_i z_i \left[D \odot H^\top\begin{pmatrix}\nabla\log\pi(x) \\ \mathbf{0}_{\tilde{d}-d}\end{pmatrix}\right]_i\right)}\right)
\end{equation*}
\end{algorithmic}
\end{algorithm}

\section{A Gibbs-style Barker algorithm}
\label{sec:barker_gibbs}

\begin{figure}[h]
    \centering
    \includegraphics[width=0.8\textwidth]{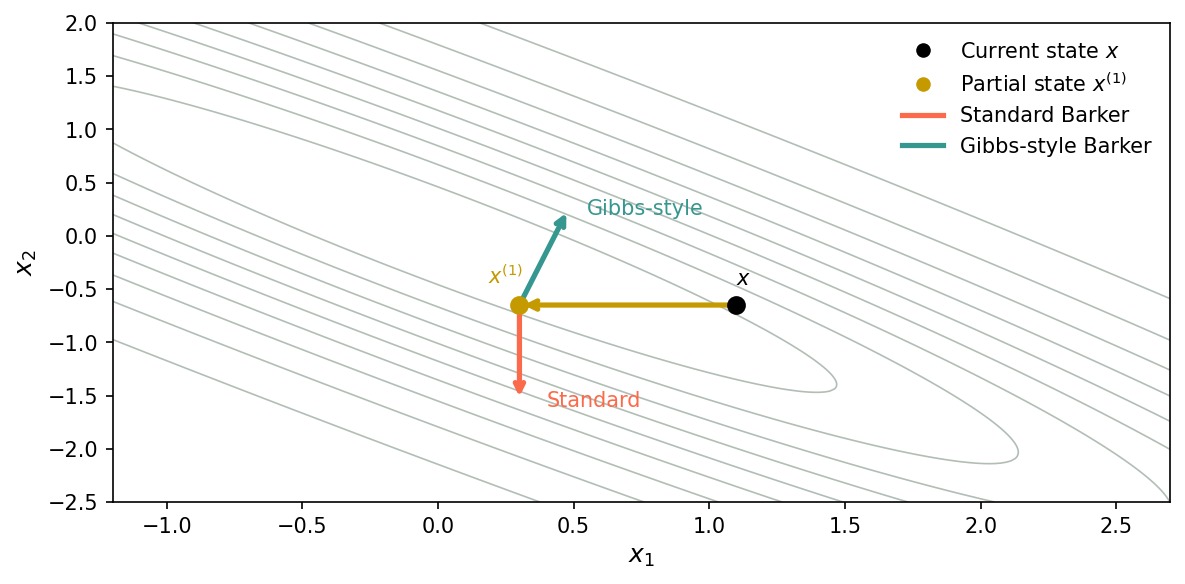}
    \caption{Standard and Gibbs-style Barker proposals on a correlated bivariate Gaussian target ($\rho =0.95$). After the first-coordinate move from the current state $x$ to the partial state $x^{[1]}$, the standard scheme proposes the second coordinate using $\partial_2\log\pi(x)$, while the Gibbs-style scheme uses $\partial_2\log\pi(x^{[1]})$.}
    \label{fig:correlated_gaussian_proposal}
\end{figure}

The standard coordinate-wise Barker proposal freezes the gradient $\nabla\log\pi(x)$ at the current state $x$ and uses it to propose all $d$ coordinate increments independently. When the target exhibits correlations or local geometric structure across coordinates, this may discard useful information, as the optimal move in each coordinate direction depends on the moves in others. To address this, we introduce another Barker variant, which we call Gibbs-style Barker, that sweeps through coordinates in a fixed order and evaluates the gradient at the partially-updated state before each coordinate proposal. More precisely, it generates a proposal using the density
\begin{equation*}
    q(y|x) \propto \prod_{i=1}^d 2 \cdot F(c_F \cdot [\nabla\log\pi(x^{[i-1]})]_i \cdot z_i) \cdot f_\sigma(z_i), \quad z_i:= y_i - x_i,
\end{equation*}
where $x^{[i-1]} := (y_1, \ldots, y_{i-1}, x_i, \ldots, x_d)$ is the $(i-1)$-th partial state, with $x^{[0]} = x$. The contribution when $i = 1$ coincides with the standard Barker proposal density for the first coordinate. For $i > 1$, however, the Gibbs-style Barker uses $[\nabla\log\pi(x^{[i-1]})]_i$ instead of $[\nabla\log\pi(x)]_i$. In particular, if the target factorizes as $\pi(x) = \prod_{i=1}^d \pi_i(x_i)$, so that $[\nabla\log\pi(x)]_i$ depends only on $x_i$ and $[\nabla\log\pi(x^{[i-1]})]_i = [\nabla\log\pi(x)]_i$, then the two proposals are the same. Any difference between the two methods therefore only manifests when the coordinates are not independent. Figure~\ref{fig:correlated_gaussian_proposal} shows this on a bivariate Gaussian with $\rho = 0.95$, where the two proposals point in different directions after the first coordinate move. Re-evaluating the gradient at the partially updated state $x^{[i-1]}$ produces a second-coordinate increment better aligned with the local elliptical geometry of the target.

\begin{algorithm}
\caption{Gibbs-style Barker}
\label{alg:gibbs_barker}
\begin{algorithmic}[1]
\Require Current state $X^{(n)}= x \in \mathbb{R}^d$, step size $\sigma$
\For{$i \in \{1, \ldots, d\}$} draw $z_i \sim f_\sigma$
\EndFor
\State Set $x^{[0]} = x$
\For{$i \in \{1, \ldots, d\}$}
    \State Draw $b_i \sim \mathrm{Rademacher}\!\left(F\!\left(c_F \cdot [\nabla\log\pi(x^{[i-1]})]_i \cdot z_i\right)\right)$
    \State Set $x^{[i]} = x^{[i-1]}$ with $x^{[i]}_i = x^{[i-1]}_i + b_i z_i$
\EndFor
\State Set $y = x^{[d]}$
\State Set $X^{(n+1)} = y$ with probability $\alpha(x, y)$, otherwise set $X^{(n+1)} = x$, where
\begin{equation*}
    \alpha(x, y) = \min\!\left(1,\; \frac{\pi(y)}{\pi(x)} \cdot \prod_{i=1}^{d} \frac{1 - F\!\left(c_F \cdot (y_i - x_i)\, [\nabla\log\pi(y^{[i-1]})]_i\right)}{F\!\left(c_F \cdot (y_i - x_i)\, [\nabla\log\pi(x^{[i-1]})]_i\right)}\right)
\end{equation*}
with $y_j^{[i-1]} = x_j$ for $j < i$ and $y_j^{[i-1]} = y_j$ for $j \ge i$.
\end{algorithmic}
\end{algorithm}

Mathematically, the Gibbs-style Barker proposal requires $d$ partial derivatives to be evaluated, and another $d$ to compute the Metropolis--Hastings filter.  This is the same as for the standard Barker proposal, meaning that the two methods should have equivalent computational cost.  The iterative manner in which these partial derivatives must be computed, however, can lead to additional computational overheads depending on the model, the programming language used and the specific style of implementation.  We illustrate this in Section \ref{sec:numerics}.  

\section{A Manifold Barker algorithm}
\label{sec:barker_manifold}

Distributions of interest that possess varying local geometry, non-uniform scaling, or correlation structures that change across the state space pose a challenge for samplers based on a constant global preconditioner. A natural remedy is to let the preconditioning matrix adapt to the local geometry, resulting in a position-dependent transformation.  In this section we devise a variant of the Barker proposal that can operate well in such a setting, based on the theory of Riemannian sampling algorithms, which are briefly reviewed below.

\subsection{Riemannian Langevin Diffusion}

Returning to the overdamped Langevin diffusion defined by (\ref{eq:langevin_diffusion}), making the preconditioning matrix state-dependent requires a corresponding adjustment to the drift so that the stationary Fokker–Planck equation still holds and $\pi$ remains invariant \parencite{xifara2014langevin}. The resulting diffusion process is governed by the stochastic differential equation
\begin{equation*}
    dX(t) = \left( \frac{1}{2} A(X(t))\, \nabla \log \pi(X(t)) + \Gamma(X(t)) \right)dt + \sqrt{A(X(t))}\, dW(t), \quad X(0) = x,
\end{equation*}
with the correction term 
\begin{equation*}
    \Gamma_i(x) = \frac{1}{2} \sum_{j=1}^{d} \frac{\partial A_{ij}(x)}{\partial x_j}.
\end{equation*}
This is sometimes called the Riemannian overdamped Langevin diffusion, following the terminology of \textcite{girolami2011riemann} who derived this SDE from a Riemannian geometric construction (see also Section 4 of \textcite{livingstone2014information}). The Euler--Maruyama discretization of this stochastic differential equation with approximate transition density $N(x + sA(x) \nabla\log\pi(x)/2 + s\Gamma(x), sA(x))$ for suitably chosen $s>0$ is the manifold MALA (MMALA) proposal distribution used by the same authors to define a Metropolis--Hastings algorithm that can navigate irregularly-shaped distributions of interest. 

The function $A: \mathbb{R}^d \to \mathbb{R}^{d \times d}$ is required to be positive definite at every $x$, but is otherwise a free design decision that is originally motivated by geometric considerations \parencite{girolami2011riemann}. A natural objective is to choose $A(x)^{-1}$ to match the local covariance structure of $\pi$. An intuitive choice is therefore the Hessian $A(x)^{-1} = -\nabla^2 \log\pi(x)$, though this may not be positive definite everywhere when $\pi$ is not log-concave. One approach to address this is to use the SoftAbs metric \parencite{betancourt2013general}, which applies a smooth transformation to each eigenvalue of the metric via $t_\alpha(\lambda_i) = \lambda_i(\coth(\alpha\lambda_i))$, where $\alpha > 0$ is a tuning parameter chosen to be as large as possible without compromising numerical stability. For large $|\lambda_i|$, this mapping approximates $|\lambda_i|$, while eigenvalues near zero are smoothly regularized to $1/\alpha$. Other variants and non-Hessian based choices are also possible \parencite{girolami2011riemann,livingstone2014information}.

Evaluating $\Gamma(x)$ requires derivatives of $A(x)$, which in turn requires third derivatives of $\log\pi(x)$ when $A(x)$ is Hessian based. This is prohibitively expensive in many settings. The simplified Manifold MALA (SMMALA) of \textcite{girolami2011riemann} drops this correction term, giving the proposal $N(x + s A(x) \nabla\log\pi(x)/2, s A(x))$ for small time step $s$, which can be understood as the Euler--Maruyama discretization of a simplified Riemannian Langevin diffusion defined by 
\begin{equation} \label{eq:simplified_riemannian_langevin_sde}
    dX(t) = \frac{1}{2} A(X(t))\, \nabla \log \pi(X(t))dt + \sqrt{A(X(t))}\, dW(t), \quad X(0)=x.
\end{equation}
In general, this simplified equation no longer has $\pi$ as an invariant distribution, but using the SMMALA proposal combined with a Metropolis–Hastings correction nonetheless recovers $\pi$ as the invariant distribution of the resulting Markov chain. Moreover, it is shown in \textcite{girolami2011riemann} that SMMALA performs comparably to MMALA on a range of targets while avoiding the third-derivative computation, leading to a more efficient algorithm in terms of effective samples produced per unit of time owing to its lower cost per iteration.

\subsection{Simplified Manifold Barker}

Applying the skew-symmetric numerical scheme of \textcite{iguchi2026skew} to the simplified Riemannian Langevin diffusion \eqref{eq:simplified_riemannian_langevin_sde} produces a manifold version of the Barker proposal. Let $L(x)$ be the Cholesky factor of $A(x)$, so that $L(x)L(x)^\top = A(x)$. The drift and diffusion coefficients of the diffusion are
\begin{equation*}
    \mu(x) = \frac12 A(x)\nabla\log\pi(x) = \frac12 L(x) L(x)^\top \nabla\log\pi(x), \qquad
\sigma(x) = \sqrt{A(x)} = L(x).
\end{equation*}
The scaled drift $\Sigma^{-1/2}(x) \mu(x)$, which determines the level of skew in each component, simplifies to
\begin{equation*}
    \Sigma^{-1/2}(x) \mu(x) = L(x)^{-1} \cdot \frac12 A(x) \nabla\log\pi(x) = \frac12 L(x)^\top \nabla\log\pi(x).
\end{equation*}
The proposal density is therefore
\begin{equation} \label{eq:smmaba_proposal}
    q(y|x) \propto \prod_{i=1}^d 2 \cdot F\!\left(c_F \cdot [L(x)^\top \nabla\log\pi(x)]_i \cdot z_i\right) \cdot f_{s^{1/2}}(z_i), \qquad z := L(x)^{-1}(y - x)
\end{equation}
where taking $F$ to be the logistic function gives $F'(0) = 1/4$ and hence $c_F = 1$.

With $F$ being the logistic function, a proposal from (\ref{eq:smmaba_proposal}) can be generated by drawing $z_i \sim f_{s^{1/2}}$ and $b_i \sim \text{Rademacher}(p_i(x, z_i))$ independently for $i = 1, \ldots, d$, with
\begin{equation*}
    p_i(x, z) = \frac1{1 + \exp\{-z_i [L(x)^\top \nabla\log\pi(x)]_i\}},
\end{equation*}
and then setting $y = x + L(x)(b \odot z)$. The resulting sampler is the Barker analogue of SMMALA, with the drift $s A(x)\nabla\log\pi(x)/2$ replaced by coordinate-wise sign flips in the preconditioned basis, whose probabilities are determined by the above equation. We refer to it as the simplified manifold Barker algorithm (SMBarker), in analogy with SMMALA.

To express $q(y|x)$ as a density in $y$, we use the change of variables $z = L(x)^{-1}(y-x)$, which has Jacobian $|L(x)|^{-1}$. Since $L(x)$ depends on the current state $x$, this Jacobian does not cancel trivially as in the constant-preconditioner case. Instead, the forward and reverse proposal contribute to the Metropolis–Hastings ratio an additional factor of $|L(x)|/|L(y)|$. The full algorithm is given in Algorithm \ref{alg:smbarker}, in which we write $f_{d ,\sigma}(z) := \prod_i f_\sigma(z_i)$ for the $d$-dimensional probability density $f_{d,\sigma}$.

\begin{algorithm}
\caption{Simplified Manifold Barker}
\label{alg:smbarker}
\begin{algorithmic}[1]
\Require Current state $X^{(n)} = x \in \mathbb{R}^d$, step size $\sigma$, position-dependent preconditioner $L(x) \in \mathbb{R}^{d \times d}$ such that $L(x)L(x)^T = A(x)$
\For{$i \in \{1, \ldots, d\}$} draw $z_i \sim f_\sigma$ 
\EndFor
\For{$i \in \{1, \ldots, d\}$} draw $b_i \sim \mathrm{Rademacher}\!\left(F\!\left(c_F \cdot [L(x)^\top \nabla\log\pi(x)]_i \cdot z_i\right)\right)$
\EndFor
\State Set $y = x + L(x)(b \odot z)$ (where $\odot$ denotes the element-wise product)
\State Set $X^{(n+1)} = y$ with probability $\alpha(x, y)$, otherwise set $X^{(n+1)} = x$, where
\begin{align*}
    \alpha(x, y) = \min\!\bigg(1,\; \frac{\pi(y)}{\pi(x)} \cdot\frac{|L(x)|}{|L(y)|}\cdot 
    & 
    \frac{f_{d,\sigma}(L(y)^{-1}L(x)(b\odot z))}{f_{d,\sigma}(b\odot z)} \cdot 
    \\
    &
    \prod_{i=1}^d \frac{1 - F\!\left(c_F \cdot [L(y)^{-1}L(x)(b\odot z)]_{i}\cdot[L(y)^{T}\nabla\log\pi(y)]_i\right)}{F\!\left(c_F \cdot b_i z_i [L(x)^\top \nabla\log\pi(x)]_i\right)}\bigg)
\end{align*}
\end{algorithmic}
\end{algorithm}

The identity $B^{-1}=B^{T}(BB^{T})^{-1}$ for real invertible matrices can be used in Step 6 of Algorithm \ref{alg:smbarker}, meaning $L(y)^{-1}L(x)=L(y)^{T}A^{-1}(y)L(x)$, which may be an easier matrix to compute in practice.  The per-iteration cost of SMBarker (as with SMMALA) is typically dominated by the Hessian evaluation and its eigendecomposition, which is required by the SoftAbs method at every iteration. We evaluate the empirical performance of SMBarker against SMMALA and other algorithms in Section \ref{sec:numerics}.

\section{Numerical Study}
\label{sec:numerics}

We present two sets of experiments. The first compares the Gibbs-style, Hadamard, and Rotated Barker variants against the standard Barker proposal on a high–dimensional Gaussian target distribution with known precision matrix and on a Bayesian logistic regression posterior using the Sonar dataset. The second compares SMBarker against MALA, Barker, and SMMALA on two target distributions with non-linear correlation structure, namely the hybrid Rosenbrock distribution and the FitzHugh--Nagumo model. For SMMALA and SMBarker, the position-dependent preconditioner is taken to be SoftAbs metric of \textcite{betancourt2013general}, meaning the Hessian $A(x)^{-1} = -\nabla^2\log\pi(x)$ regularized by the SoftAbs function as described in Section \ref{sec:barker_manifold}. All other methods use an identity preconditioner $L = I$.  All algorithms are implemented in Python to ensure a fair comparison of computational costs.

Sampling efficiency is measured using effective sample size (ESS), which approximates the number of independent draws from the target distribution that would give the same Monte Carlo precision as the correlated chain \parencite{roy2020convergence}. For each run, the effective sample size is computed separately for each coordinate and the minimum across coordinates is taken as a summary. We report this minimum ESS, ESS per iteration, and ESS per CPU second. Each metric is averaged across independent runs and computed using post-burn-in iterations only.

Step sizes are tuned adaptively during burn-in iterations using Algorithm 4 of \textcite{andrieu2008tutorial} and held fixed thereafter, with all performance metrics computed on the main iterations only. The target acceptance probability is set to $0.574$ for all algorithms following the asymptotic optimality results for MALA and the Barker proposal \parencite{roberts1998optimal,vogrinc2023optimal}. 

\subsection{High-dimensional Gaussian}
\label{subsec:numerics_gaussian}

In the first experiment the target distributions are zero-mean multivariate normal distributions with known precision matrix $A$, meaning $\pi(x) \propto \exp\{- x^\top A x/2 \}$.  The matrix $A$ is drawn from a Wishart distribution with identity scale matrix and $d$ degrees of freedom. The same $A$ is used for all samplers within a trial, but $A$ is regenerated independently across trials. We consider $8$ values of $d$, log-spaced between $10$ and $122$, with $30$ trials per dimension and $200{,}000$ main iterations per trial (after $5{,}000$ burn-in iterations). This allows us to assess sampler performance on strongly correlated targets across a range of dimensions. The results are given in Table~\ref{tab:exp_mvn_summary}.

\begin{table}[ht]
\centering
\begin{tabular}{ll rrrr}
\hline
$d$ & Sampler & Min ESS & Min ESS/s & Min ESS/iter & CPU (s) \\
\hline
10 & Barker & 129.98 & \textbf{15.718} & 6.50e-04 & \textbf{8.3} \\
 & Gibbs-style Barker & \textbf{177.24} & 5.623 & \textbf{8.86e-04} & 31.1 \\
 & Hadamard Barker & 126.77 & 7.391 & 6.34e-04 & 17.4 \\
 & Rotated Barker & 125.60 & 4.872 & 6.28e-04 & 25.6 \\
\hline
14 & Barker & 62.94 & \textbf{7.301} & 3.15e-04 & \textbf{8.8} \\
 & Gibbs-style Barker & \textbf{88.88} & 2.166 & \textbf{4.44e-04} & 41.1 \\
 & Hadamard Barker & 59.06 & 3.349 & 2.95e-04 & 17.4 \\
 & Rotated Barker & 64.15 & 2.144 & 3.21e-04 & 29.8 \\
\hline
20 & Barker & 25.65 & \textbf{2.883} & 1.28e-04 & \textbf{9.1} \\
 & Gibbs-style Barker & \textbf{36.12} & 0.650 & \textbf{1.81e-04} & 55.8 \\
 & Hadamard Barker & 30.08 & 1.653 & 1.50e-04 & 18.3 \\
 & Rotated Barker & 26.24 & 0.764 & 1.31e-04 & 34.3 \\
\hline
29 & Barker & 9.50 & \textbf{1.045} & 4.75e-05 & \textbf{9.5} \\
 & Gibbs-style Barker & \textbf{11.98} & 0.156 & \textbf{5.99e-05} & 78.2 \\
 & Hadamard Barker & 10.80 & 0.619 & 5.40e-05 & 18.5 \\
 & Rotated Barker & 9.47 & 0.228 & 4.73e-05 & 42.0 \\
\hline
41 & Barker & \textbf{7.82} & \textbf{0.818} & \textbf{3.91e-05} & \textbf{10.0} \\
 & Gibbs-style Barker & 7.65 & 0.072 & 3.82e-05 & 108.7 \\
 & Hadamard Barker & 7.11 & 0.386 & 3.55e-05 & 19.9 \\
 & Rotated Barker & 6.50 & 0.121 & 3.25e-05 & 55.3 \\
\hline
59 & Barker & 3.98 & \textbf{0.370} & 1.99e-05 & \textbf{11.1} \\
 & Gibbs-style Barker & \textbf{4.26} & 0.027 & \textbf{2.13e-05} & 154.1 \\
 & Hadamard Barker & 3.65 & 0.167 & 1.83e-05 & 20.9 \\
 & Rotated Barker & 4.23 & 0.022 & 2.12e-05 & 190.5 \\
\hline
85 & Barker & 2.88 & \textbf{0.234} & 1.44e-05 & \textbf{12.7} \\
 & Gibbs-style Barker & 3.17 & 0.015 & 1.59e-05 & 218.9 \\
 & Hadamard Barker & \textbf{3.27} & 0.092 & \textbf{1.64e-05} & 49.9 \\
 & Rotated Barker & 3.11 & 0.010 & 1.56e-05 & 317.0 \\
\hline
122 & Barker & 2.14 & \textbf{0.092} & 1.07e-05 & \textbf{30.4} \\
 & Gibbs-style Barker & \textbf{2.40} & 0.004 & \textbf{1.20e-05} & 637.4 \\
 & Hadamard Barker & 2.18 & 0.050 & 1.09e-05 & 55.3 \\
 & Rotated Barker & 1.89 & 0.003 & 0.95e-05 & 551.3 \\
\hline
\end{tabular}
\caption{Performance of Barker, Gibbs-style Barker, Hadamard Barker, and Rotated Barker on correlated Gaussian targets, measured by minimum ESS, ESS per second, ESS per iteration, and CPU time (seconds), across dimensions $d \in \{10, 14,20, 29, 41, 59, 85, 122\}$. Results are averaged over $30$ runs, each producing $200{,}000$ samples. Bold entries indicate the best-performing method per metric within each dimension.}
\label{tab:exp_mvn_summary}
\end{table}

\begin{figure}[h]
    \centering
    \includegraphics[width=\textwidth]{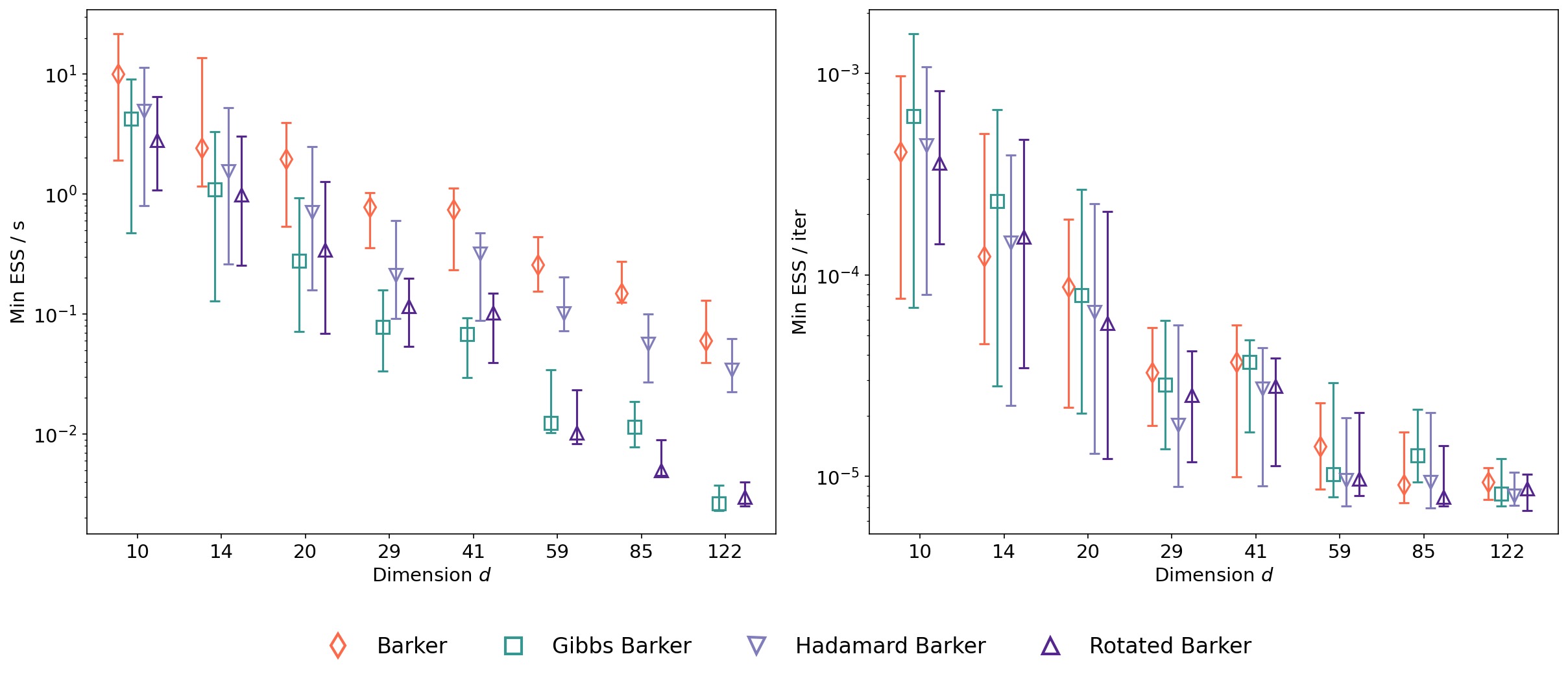}
    \caption{Minimum ESS per second (left) and per iteration (right) of Barker, Gibbs-style Barker, Hadamard Barker, and Rotated Barker on correlated Gaussian targets, across dimensions $d \in \{10, 14,20, 29, 41, 59, 85, 122\}$. Markers indicate the median and whiskers show the interquartile range across $30$ trials, each producing $200{,}000$ samples. Both axes are on a logarithmic scale.}
    \label{fig:exp_mvn_whisker_2panel}
\end{figure}

Standard Barker achieves the best time-normalized efficiency across all dimensions, despite not having the best raw ESS at any dimension other than $d=41$. Gibbs-style Barker has the highest raw ESS at $6$ of the $8$ considered dimensions, although its CPU cost is by far the largest, between $3$ and $21$ times slower than standard Barker depending on dimension. Of the three variants, Hadamard Barker offers the best balance between statistical and computational efficiency, achieving ESS competitive with standard Barker while incurring lower CPU cost than Gibbs or Rotated Barker. By contrast, Rotated Barker scales worst with dimension. Its CPU time grows from comparable to Hadamard Barker at low $d$ to nearly $10$ times higher at $d=122$, and it has the lowest ESS/s of the four methods at $d \ge 59$.

Two implementation caveats apply. First, Gibbs-style Barker's CPU cost is inflated by the implementation-level overhead of repeatedly evaluating partial derivatives at each partial state, which a lower-level implementation would reduce. In particular, the CPU time increases by a factor of around $2.9$ between $d=85$ and $d=122$, exceeding what algorithmic complexity alone would predict and suggesting that interpreter overhead becomes increasingly dominant at higher dimensions. Second, Hadamard Barker's $O(d^2)$ matrix-vector product could be replaced by the fast Walsh--Hadamard transform at $O(d \log d)$ cost.

\subsection{Logistic regression with the Sonar dataset}
\label{subsec:numerics_logistic}

We perform Bayesian inference for a logistic regression model on the Sonar dataset \parencite{gorman1988analysis}, which has $208$ observations and $60$ covariates. The model includes an intercept, giving a $61$-dimensional parameter space. A multivariate standard Normal prior is placed on the parameter vector, and all covariates are normalized to have zero mean and unit variance. Each sampler is run for $200{,}000$ iterations across $10$ independent trials, and results are reported in Table~\ref{tab:exp_sonar_summary}.


Gibbs-style Barker again dominates the raw ESS comparison, achieving roughly $1.6$ times the value of the next-best method. Hadamard Barker remains the most practically competitive variant, slightly behind standard Barker in ESS/s while delivering $13\%$ more on raw ESS. Rotated Barker is now effectively tied with standard Barker on raw ESS, but its order-of-magnitude higher CPU cost again places it last on ESS/s. The implementation caveats noted in the previous section apply here as well. 

\begin{table}[ht]
\centering
\begin{tabular}{l rrrr}
\hline
Sampler & Min ESS & Min ESS/s & Min ESS/iter & CPU (s) \\
\hline
Barker          & 375.4 & \textbf{19.267} & 1.88e-03 & \textbf{19.54} \\
Gibbs-style Barker    & \textbf{662.1} & 1.624 & \textbf{3.31e-03} & 407.87 \\
Hadamard Barker & 425.2 & 14.546 & 2.130e-03 & 29.24 \\
Rotated Barker  & 377.8 & 1.880 & 1.89e-03 & 201.05 \\
\hline
\end{tabular}
\caption{Performance of Barker, Gibbs-style Barker, Hadamard Barker, and Rotated Barker on the Sonar dataset, measured by minimum ESS, ESS per second, ESS per iteration, and CPU time (seconds). Results are averaged over $10$ trials, each producing $200{,}000$ samples. Bold entries indicate the best-performing method per metric.}
\label{tab:exp_sonar_summary}
\end{table}

\subsection{Hybrid Rosenbrock distribution}
\label{subsec:numerics_rosenbrock}

The hybrid Rosenbrock distribution \parencite{pagani2022n} has density
\begin{equation*}
    \pi(x) \propto \exp\left\{-a(x_1 - \mu)^2 - \sum_{j=1}^{n_2}\sum_{i=2}^{n_1} b\left(x_{j,i} - x_{j,i-1}^2\right)^2\right\},
\end{equation*}
with dimension $d = (n_1 - 1)n_2 + 1$. Its defining feature is a narrow curved ridge in every two-dimensional marginal, which makes exploration challenging for samplers with isotropic proposals. We use the five-dimensional version with $n_1 = 3$, $n_2 = 2$, $\mu = 1$, $a = 0.05$ throughout, and vary the parameter $b$ across experiments.

The hybrid Rosenbrock distribution is sufficiently challenging that none of the samplers achieve clear convergence within our compute budget, making ESS-based metrics unreliable. We instead report the Kolmogorov--Smirnov (KS) and Anderson--Darling (AD) distances between the marginal empirical distribution of each coordinate and the corresponding reference marginal, taking the maximum across coordinates as a worst-case summary. The KS distance is the supremum of the absolute difference between two cumulative distribution functions, and the AD distance is a weighted variant that assigns more weight to discrepancies in the tails of the distribution, where exploration is most difficult. 

For this target distribution the asymptotic optimal acceptance rate of $0.574$ is not appropriate for all considered samplers, and step sizes are instead tuned to minimize KS and AD distances at $b=5$ across a range of target acceptance probabilities $\{0.1, 0.25, 0.4, 0.55, 0.7, 0.85\}$. The tuning behavior varied substantially between samplers. SMBarker consistently achieved its target acceptance probability very quickly during the $100{,}000$ burn-in iterations, while SMMALA tended to undershoot, yielding true acceptance rates lower than the specified target. MALA and Barker exhibited stable performance across the considered range of acceptance probabilities, suggesting that their performance is relatively insensitive to fine-tuning on this target. In contrast, SMMALA and SMBarker showed sharp increases in both KS and AD distances at higher acceptance probabilities. One explanation is that the position-dependent metric enables more ambitious proposals, which improve exploration when the step size is well-tuned but lead to instability otherwise. Following the recommendation of \textcite{pagani2022n} and our tuning experiment, we set the target acceptance probability to $0.25$ for SMMALA and $0.1$ for SMBarker, while MALA and Barker retain the theoretically optimal value of $0.574$. Step sizes were not re-tuned across other values of $b$, as the tuning results of \textcite{pagani2022n} suggest that the optimal step size varies little with $b$ relative to the variability of the tuning procedure itself. To ensure fair comparison, the number of analyzed iterations was truncated for each sampler, so that all methods correspond to the same compute budget, determined by the fastest sampler. Results are summarized in Table~\ref{tab:exp_rosenbrock_summary}.

\begin{table}[ht]
\centering
\begin{tabular}{ll rrr}
\hline
$b$ & Sampler & Adjusted Iterations & Mean Max KS & Mean Max AD \\
\hline
0.005 & Barker    & 4,792,346 & 0.263 & 6050.8 \\
      & MALA    & 5,000,000 & 0.337 & 9920.2 \\
      & SMBarker  & 1,252,578 & \textbf{0.027} & \textbf{25.6} \\
      & SMMALA  & 1,150,406 & 0.905 & 1{,}018{,}499.0 \\
\hline
0.05  & Barker    & 4,612,610 & 0.357 & 9750.4 \\
      & MALA    & 5,000,000 & 0.381 & 12453.7 \\
      & SMBarker  & 1,229,642 & \textbf{0.049} & \textbf{96.6} \\
      & SMMALA  & 1,125,325 & 0.905 & 1{,}023{,}391.0 \\
\hline
0.5   & Barker    & 4,704,230 & 0.387 & 12235.1 \\
      & MALA    & 5,000,000 & 0.487 & 20307.9 \\
      & SMBarker  & 1,228,284 & \textbf{0.093} & \textbf{481.4} \\
      & SMMALA  & 1,132,648 & 0.852 & 799{,}739.3 \\
\hline
5     & Barker    & 4,832,702 & 0.485 & 19972.6 \\
      & MALA    & 5,000,000 & 0.575 & 27916.8 \\
      & SMBarker  & 1,248,584 & \textbf{0.121} & \textbf{892.3} \\
      & SMMALA  & 1,150,424 & 0.861 & 828{,}241.2 \\
\hline
50    & Barker    & 4,660,017 & 0.557 & 25142.7 \\
      & MALA    & 5,000,000 & 0.640 & 34056.0 \\
      & SMBarker  & 1,240,255 & \textbf{0.224} & \textbf{2924.0} \\
      & SMMALA  & 1,141,924 & 0.878 & 903{,}527.9 \\
\hline
\end{tabular}
\caption{Performance of Barker, MALA, SMBarker, and SMMALA across values of $b$, measured by maximum KS and AD distances evaluated on adjusted iterations. Results are averaged over $30$ runs.  Bold entries indicate the best-performing method per metric for each $b$ value.}
\label{tab:exp_rosenbrock_summary}
\end{table}

\begin{figure}[h]
    \centering
    \includegraphics[width=\textwidth]{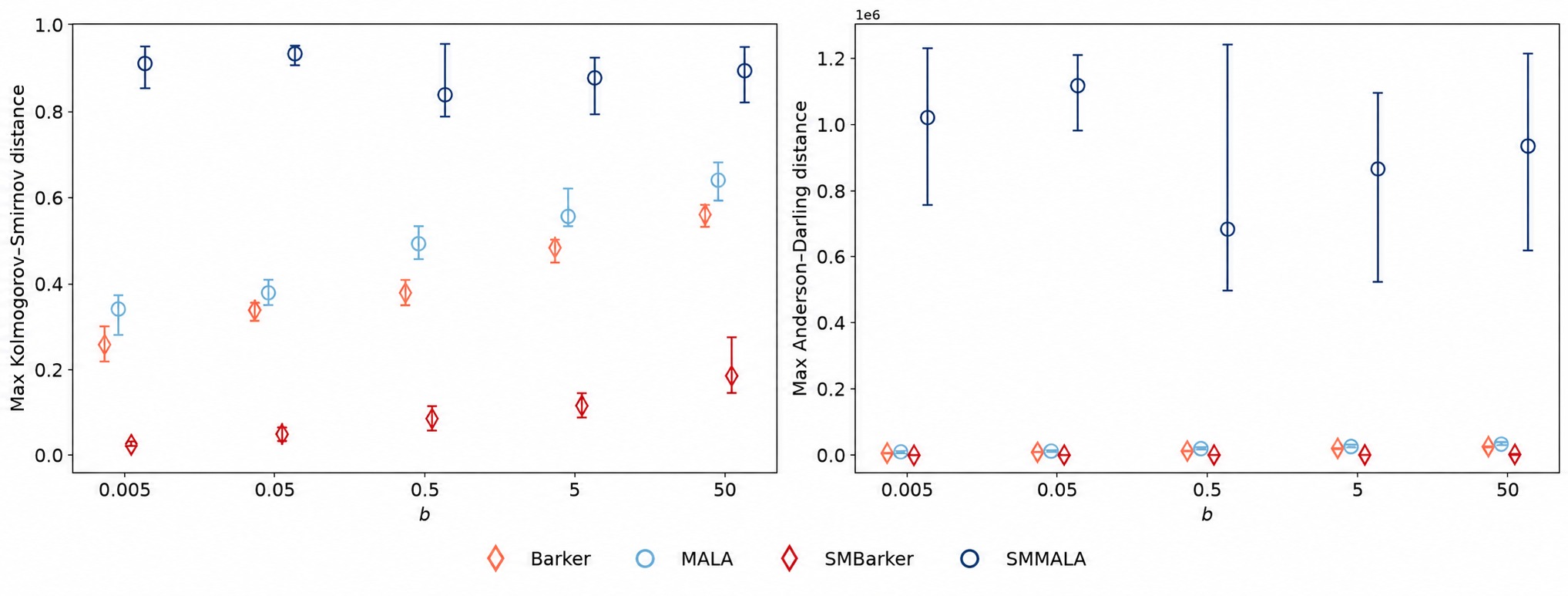}
    \caption{Maximum KS distance (left) and maximum AD distance (right) between the marginal empirical distributions and the corresponding reference marginals on the hybrid Rosenbrock target distribution, taken across coordinates. Markers indicate the median and whiskers show the interquartile range across $30$ trials. } 
    \label{fig:exp_rosenbrock_median_iqr}
\end{figure}


SMBarker achieved the lowest KS and AD distances for all different values of $b$. By contrast the performance of SMMALA deteriorated as $b$ was reduced.  One explanation for these outcomes is the instability of the position-dependent preconditioner based on the inverse Hessian, though the failure modes differ between the two methods. When $b$ is small, the second derivatives of $\log\pi$ approach zero, causing the Hessian to become near-singular. 
The enhanced stability of SMBarker suggests that the robustness advantage of the Barker proposal over MALA for the constant preconditioner case, documented in \textcite{livingstone2022barker}, also extends to the position-dependent setting. 
Among the constant-preconditioner methods, Barker outperformed MALA at every value of $b$ in terms of both KS and AD distances, consistent with the same robustness property.


\subsection{Fitzhugh--Nagumo model}
\label{subsec:numerics_fitzhugh}

Finally, we consider posterior inference for the three parameters of the FitzHugh--Nagumo differential equations $\dot{V} = c\left(V - V^3/3 + R\right)$ and $\dot{R} = -(V - a + bR)/c$ \parencite{ramsay2007parameter}. Following the setup of \textcite{girolami2011riemann}, we simulated $200$ data points from this system over $t \in [0, 20]$, using parameters $a=0.2$, $b=0.2$, $c=3$ and initial conditions $V(0) = -1$, $R(0) = 1$, and added independent Gaussian noise with standard deviation $0.5$ to both state variables. Ten such datasets were generated with different random seeds, and each sampler was run $10$ times per dataset, collecting $5{,}000$ posterior samples. As in \textcite{girolami2011riemann}, all chains were initialized at the true mode, thereby sidestepping the exploration challenges associated with multimodal target distributions.

The per-iteration computational cost for this target distribution was dominated by the numerical integration of the FitzHugh--Nagumo system required to evaluate the likelihood, so all four samplers had essentially the same CPU time per run.  Time-normalized ESS metrics are therefore less informative about the methodological differences, so we instead focus the comparison on raw ESS values.

\begin{table}[ht]
\centering
\begin{tabular}{l rrr r}
\hline
Sampler & ESS $a$ & ESS $b$ & ESS $c$ & CPU/Min ESS \\
\hline
Barker  & $152.1\ (23.3)$ & $33.3\ (10.6)$ & $86.6\ (49.9)$  & $5.857$ \\
MALA    & $121.4\ (23.6)$ & $21.4\ (4.8)$  & $65.5\ (40.5)$  & $7.081$ \\
SMBarker & $576.0\ (84.8)$ & $348.9\ (43.6)$ & $474.9\ (84.7)$ & $0.366$ \\
SMBarker ($0.4$) & $607.3\ (99.0)$ & $407.7\ (55.2)$ & $530.8\ (87.9)$ & \textbf{0.295} \\
SMMALA  & $\textbf{850.4}\ (155.5)$ & $\textbf{445.0}\ (45.5)$ & $\textbf{652.0}\ (153.2)$ & $0.304$ \\  
\hline
\end{tabular}
\caption{Performance of Barker, MALA, SMBarker, and SMMALA on the FitzHugh--Nagumo posterior, measured by ESS for each parameter with standard deviation in parentheses. Results are averaged over $100$ runs per sampler ($10$ datasets, $10$ runs per dataset, $5{,}000$ posterior samples per run). Bold entries indicate the best value per column.  SMBarker ($0.4$) denotes a version with target acceptance rate $0.4$ rather than $0.57$.}
\label{tab:fhn}
\end{table}

\begin{figure}[h]
    \centering
    \includegraphics[width=\textwidth]{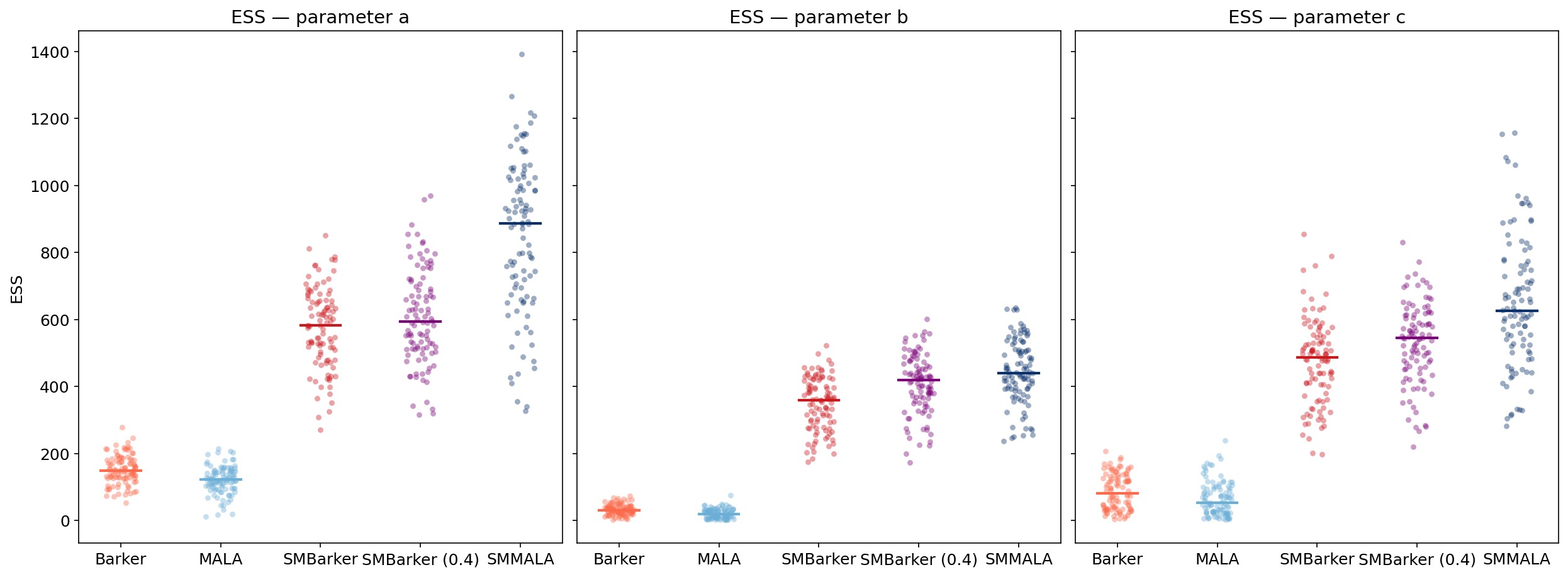}
    \caption{Effective sample sizes for each parameter $(a, b, c)$ of the FitzHugh--Nagumo model. Each point represents a single run and horizontal bars show the mean across $100$ runs per sampler.}
    \label{fig:exp_fhn_ess_strip}
\end{figure}

In terms of raw ESS, both locally-preconditioned methods substantially outperformed the constant-preconditioner methods, with the gains ranging from roughly four times to over ten times as many effective samples across the three parameters. Among the two locally-preconditioned methods, SMMALA achieved the highest raw ESS, with SMBarker reaching about $80\%$ of SMMALA's value on the worst parameter $b$. The observed relative performance of SMBarker and SMMALA on this target distribution is consistent with the experimental design, which favors methods that aggressively exploit the reliable local curvature information. SMMALA's Gaussian shift along the preconditioned gradient direction does this more efficiently than SMBarker's coordinate-wise sign flip. 

\section{Discussion}

Skew-symmetric distributions have proven useful in many application areas.  We have highlighted their effectiveness in designing Markov chain Monte Carlo sampling algorithms.  The justification for their use can be understood in two different ways, as detailed in Sections \ref{subsec:motivation_langevin}-\ref{subsec:motivation_lb}.  The first is derived from a more general application of skew-symmetric distributions to approximate the transition densities of stochastic differential equations for numerical simulation purposes.  The second is via a newly studied class of Markov jump processes for which skew-symmetric distributions are often a good approximation to the jump kernel.  The resulting MCMC algorithm has some connection to the work of physicist AA Barker \parencite{barker1965monte}, and has therefore been given the name `Barker proposal' when used as a proposal kernel within a Metropolis--Hastings algorithm.

Here we present a full description of the preconditioned Barker proposal on a general $d$-dimensional Euclidean space, and review its motivation.  We then consider three natural ways to modify the algorithm and assess the relative benefits of each.  First we study coordinate-free variants of the algorithm, which rely on the introduction of an orthogonal mixture of skew-symmetric distributions.  We then consider a variant in which the degree of skew for the $i$th component when $i > 1$ depends on an innovation in the first $(i-1)$ components, meaning that the sampling of a $d$-dimensional vector from this so-called Gibbs-style Barker proposal must be done in an iterative manner (similar in spirit to the Gibbs sampler).  Finally we construct a version of the Barker proposal that exploits local geometry of the target distribution, leading to a Riemannian manifold sampler in the spirit of \textcite{girolami2011riemann}.

Some natural conclusions can be drawn from the work offered here.  The standard Barker proposal is not coordinate-invariant according to the definition of Section \ref{sec:barker_coordinatefree}, but the degree to which performance depends on the coordinate basis appears to be mild in comparison to some other algorithms (such as the Gibbs sampler).  This is evidenced by the numerical results of Sections \ref{subsec:numerics_gaussian} and \ref{subsec:numerics_logistic}, in which the raw ESSs of the standard Barker proposal and coordinate-free alternatives are consistently comparable.  When the additional costs associated with generating new bases at each iteration are taken into account, then the coordinate-free approach does not appear to offer any benefits.  Iterative computation of the degree of skew in each coordinate as in the Gibbs-style Barker algorithm does, however, appear to be advantageous in terms of raw effective sample sizes, with a factor of around 1.8 increase demonstrated for the logistic regression example using the Sonar dataset in Section \ref{subsec:numerics_logistic}.  The numerical challenge posed by iteratively evaluating the gradient seems to limit time-normalized performance, however.  In cases such as logistic regression, where each partial derivative requires looping through the entire dataset to re-evaluate quantities, then these numerical difficulties may be unavoidable without some level of approximation.  For other examples like the Gaussian, however, there is some evidence offered in the appendix that implementation in a compiled programming language such as C++ can make the sampling costs comparable between the Gibbs-style and standard Barker approaches.  The simplified manifold Barker algorithm appears to offer benefits over the simplified manifold MALA approach in terms of ease of tuning and robustness, sometimes at the cost of reduced efficiency.  This is in line with typical findings comparing the standard Barker algorithm with Gaussian symmetric noise distribution to MALA.  In that case the loss in efficiency can typically be reduced to zero by replacing the Gaussian with a bimodal symmetric density, as shown in \textcite{vogrinc2023optimal}.  The extra degree of mismatch that this causes within the acceptance rate of simplified manifold Barker, however, seems to limit the degree to which similar performance gains can be made in the manifold case.

\section*{Acknowledgments}
SL would like to thank Abdul-Lateef Haji-Ali for inspiring the development of the Gibbs-style Barker algorithm. 
PS is funded by Medical Research Council grant UKRI332.

\section*{Disclosure statement}
No potential conflict of interest was reported by the authors.

\section*{Declaration of generative AI use}
ChatGPT-5.6 Sol and Claude Sonnet were used to help write computer code for the numerical experiments and to proofread the manuscript for typographical errors and to meet journal formatting guidelines.

\section*{Appendix: Speed tests for Gibbs-style Barker}

We include a short experiment comparing the run time in three different programming languages to compute a $1000$-dimensional dense Gaussian gradient vector $\nabla g(x)$ at state $x$, a random sample from the same Gaussian with full precision matrix, and then perform the update $x + h\nabla g(x)$ for some fixed scalar $h > 0$ (Full update), compared with the iterative computation of $\partial g(x^{[i-1]})/\partial x_i$ and the update $y_i = x_i + h\partial g(x^{[i-1]})/\partial x_i$ for $ i \in \{1,...,1000\}$ where $x^{[i-1]} := (y_1,...,y_{i-1},x_i,...,x_{1000})$ (Staged update).  The results of Table \ref{tab:speed_comparison} suggest that in a compiled language such as C++ the run times are comparable, whereas in an interpreted language such as Python or R the staged update, which mimics the Gibbs-style Barker algorithm, takes significantly longer. Similar experiments with $g$ taken as the log-posterior of the logistic regression model using the Sonar dataset did not, however, give the same results, with the Gibbs-style update still taking significantly longer in C++.

\begin{table}[ht]
    \begin{center}
        \begin{tabular}{lrrr}
        \hline
        Language & Full update (s) & Staged update (s) & Ratio \\
        \hline
        Python & 0.702808 & 1.363250 & 1.940 \\
        C++ & 0.841235 & 0.843661 & 1.003 \\
        R & 0.680000 & 12.493000 & 18.372 \\
        \hline
        \end{tabular}
    \end{center}
    \caption{\label{tab:speed_comparison} Comparison of time to compute 1000 repetitions of a gradient-based update for a Gaussian target distribution in 1000 dimensions using three different programming languages.  Full update mimics the standard Barker proposal, staged update mimics the Gibbs-style Barker algorithm.}
\end{table}

\printbibliography

@article{green2015bayesian,
  title={Bayesian computation: a summary of the current state, and samples backwards and forwards},
  author={Green, Peter J and {\L}atuszy{\'n}ski, Krzysztof and Pereyra, Marcelo and Robert, Christian P},
  journal={Statistics and Computing},
  volume={25},
  number={4},
  pages={835--862},
  year={2015},
  publisher={Springer}
}

@article{vstrumbelj2024past,
  title={Past, present and future of software for Bayesian inference},
  author={{\v{S}}trumbelj, Erik and Bouchard-C{\^o}t{\'e}, Alexandre and Corander, Jukka and Gelman, Andrew and Rue, H{\aa}vard and Murray, Lawrence and Pesonen, Henri and Plummer, Martyn and Vehtari, Aki},
  journal={Statistical Science},
  volume={39},
  number={1},
  pages={46--61},
  year={2024},
  publisher={Institute of mathematical statistics}
}

@article{livingstone2022barker,
  title={The Barker proposal: Combining robustness and efficiency in gradient-based MCMC},
  author={Livingstone, Samuel and Zanella, Giacomo},
  journal={Journal of the Royal Statistical Society Series B: Statistical Methodology},
  volume={84},
  number={2},
  pages={496--523},
  year={2022},
  publisher={Oxford University Press}
}

@article{iguchi2026skew,
  title={Skew-symmetric schemes for stochastic differential equations with non-Lipschitz drift: an unadjusted Barker algorithm},
  author={Iguchi, Yuga and Livingstone, Samuel and N{\"u}sken, Nikolas and Vasdekis, Giorgos and Zhang, Rui-Yang},
  journal={IMA Journal of Numerical Analysis},
  pages={draf147},
  year={2026},
  publisher={Oxford University Press}
}

@article{iguchi2026pathwise,
  title={Pathwise skew-symmetric discretisation for SDEs with superlinear drift},
  author={Iguchi, Yuga and Livingstone, Samuel and Vasdekis, Giorgos and Zhang, Rui-Yang},
  journal={arXiv preprint arXiv:2607.17735},
  year={2026}
}

@article{power2019accelerated,
  title={Accelerated sampling on discrete spaces with non-reversible Markov processes},
  author={Power, Samuel and Goldman, Jacob Vorstrup},
  journal={arXiv preprint arXiv:1912.04681},
  year={2019}
}

@article{livingstone2025foundations,
  title={Foundations of locally-balanced Markov processes},
  author={Livingstone, Samuel and Vasdekis, Giorgos and Zanella, Giacomo},
  journal={arXiv preprint arXiv:2504.13322},
  year={2025}
}

@article{ailon2009fast,
  title={The fast Johnson--Lindenstrauss transform and approximate nearest neighbors},
  author={Ailon, Nir and Chazelle, Bernard},
  journal={SIAM Journal on computing},
  volume={39},
  number={1},
  pages={302--322},
  year={2009},
  publisher={SIAM}
}

@article{tropp2011improved,
  title={Improved analysis of the subsampled randomized Hadamard transform},
  author={Tropp, Joel A},
  journal={Advances in Adaptive Data Analysis},
  volume={3},
  number={01n02},
  pages={115--126},
  year={2011},
  publisher={World Scientific}
}

@article{xifara2014langevin,
  title={Langevin diffusions and the Metropolis-adjusted Langevin algorithm},
  author={Xifara, Tatiana and Sherlock, Chris and Livingstone, Samuel and Byrne, Simon and Girolami, Mark},
  journal={Statistics \& Probability Letters},
  volume={91},
  pages={14--19},
  year={2014},
  publisher={Elsevier}
}

@article{girolami2011riemann,
  title={Riemann manifold langevin and hamiltonian monte carlo methods},
  author={Girolami, Mark and Calderhead, Ben},
  journal={Journal of the Royal Statistical Society Series B: Statistical Methodology},
  volume={73},
  number={2},
  pages={123--214},
  year={2011},
  publisher={Oxford University Press}
}

@article{sherlock2010random,
  title={The random walk Metropolis: Linking theory and practice through a case study},
  author={Sherlock, Chris and Fearnhead, Paul and Roberts, Gareth O},
  year={2010}
}

@article{vogrinc2023optimal,
  title={Optimal design of the Barker proposal and other locally balanced Metropolis--Hastings algorithms},
  author={Vogrinc, Jure and Livingstone, Samuel and Zanella, Giacomo},
  journal={Biometrika},
  volume={110},
  number={3},
  pages={579--595},
  year={2023},
  publisher={Oxford University Press}
}

@inproceedings{hird2020fresh,
  title={A fresh take on ‘Barker dynamics’ for MCMC},
  author={Hird, Max and Livingstone, Samuel and Zanella, Giacomo},
  booktitle={International Conference on Monte Carlo and Quasi-Monte Carlo Methods in Scientific Computing},
  pages={169--184},
  year={2020},
  organization={Springer}
}

@article{arellano2006unification,
  title={On the unification of families of skew-normal distributions},
  author={Arellano-Valle, Reinaldo B and Azzalini, Adelchi},
  journal={Scandinavian Journal of Statistics},
  volume={33},
  number={3},
  pages={561--574},
  year={2006},
  publisher={Wiley Online Library}
}

@book{azzalini2013skew,
  title={The skew-normal and related families},
  author={Azzalini, Adelchi},
  volume={3},
  year={2013},
  publisher={Cambridge University Press}
}

@article{barker1965monte,
  title={Monte carlo calculations of the radial distribution functions for a proton? electron plasma},
  author={Barker, Anthony Alfred},
  journal={Australian Journal of Physics},
  volume={18},
  number={2},
  pages={119--134},
  year={1965},
  publisher={CSIRO Publishing}
}

@article{banterle2015accelerating,
  title={Accelerating Metropolis-Hastings algorithms by delayed acceptance},
  author={Banterle, Marco and Grazian, Clara and Lee, Anthony and Robert, Christian P},
  journal={arXiv preprint arXiv:1503.00996},
  year={2015}
}

@article{kloeden1989survey,
  title={A survey of numerical methods for stochastic differential equations},
  author={Kloeden, Peter E and Platen, Eckhard},
  journal={Stochastic Hydrology and Hydraulics},
  volume={3},
  number={3},
  pages={155--178},
  year={1989},
  publisher={Springer}
}

@article{roberts1996exponential,
  title={Exponential convergence of Langevin distributions and their discrete approximations},
  author={Roberts, Gareth O and Tweedie, Richard L},
  year={1996}
}

@book{milstein2004stochastic,
  title={Stochastic numerics for mathematical physics},
  author={Milstein, Grigori N and Tretyakov, Michael V},
  volume={39},
  year={2004},
  publisher={Springer}
}

@article{roberts1997updating,
  title={Updating schemes, correlation structure, blocking and parameterization for the Gibbs sampler},
  author={Roberts, Gareth O and Sahu, Sujit K},
  journal={Journal of the Royal Statistical Society Series B: Statistical Methodology},
  volume={59},
  number={2},
  pages={291--317},
  year={1997},
  publisher={Oxford University Press}
}

@article{grazzi2026randomized,
  title={On randomized step sizes in Metropolis-Hastings algorithms},
  author={Grazzi, Sebastiano and Livingstone, Samuel and Riou-Durand, Lionel},
  journal={arXiv preprint arXiv:2601.19710},
  year={2026}
}

@article{titsias2018auxiliary,
  title={Auxiliary gradient-based sampling algorithms},
  author={Titsias, Michalis K and Papaspiliopoulos, Omiros},
  journal={Journal of the Royal Statistical Society Series B: Statistical Methodology},
  volume={80},
  number={4},
  pages={749--767},
  year={2018},
  publisher={Oxford University Press}
}

@inproceedings{welling2011bayesian,
  title={Bayesian learning via stochastic gradient Langevin dynamics},
  author={Welling, Max and Teh, Yee W},
  booktitle={Proceedings of the 28th international conference on machine learning (ICML-11)},
  pages={681--688},
  year={2011}
}

@article{durmus2019high,
  title={High-dimensional Bayesian inference via the unadjusted Langevin algorithm},
  author={Durmus, Alain and Moulines, {\'E}ric},
  journal={Bernoulli},
  volume={25},
  number={4A},
  pages={2854--2882},
  year={2019}
}

@article{mezzadri2007generate,
  title={How to generate random matrices from the classical compact groups},
  author={Mezzadri, Francesco},
  journal={Notices of the American Mathematical Society},
  volume={54},
  number={5},
  pages={592--604},
  year={2007},
  publisher={American Mathematical Society}
}

@article{livingstone2014information,
  title={Information-geometric Markov chain Monte Carlo methods using diffusions},
  author={Livingstone, Samuel and Girolami, Mark},
  journal={Entropy},
  volume={16},
  number={6},
  pages={3074--3102},
  year={2014},
  publisher={MDPI}
}

@inproceedings{betancourt2013general,
  title={A general metric for Riemannian manifold Hamiltonian Monte Carlo},
  author={Betancourt, Michael},
  booktitle={International conference on geometric science of information},
  pages={327--334},
  year={2013},
  organization={Springer}
}

@article{roy2020convergence,
  title={Convergence diagnostics for markov chain monte carlo},
  author={Roy, Vivekananda},
  journal={Annual Review of Statistics and Its Application},
  volume={7},
  number={1},
  pages={387--412},
  year={2020},
  publisher={Annual Reviews}
}

@article{andrieu2008tutorial,
  title={A tutorial on adaptive MCMC},
  author={Andrieu, Christophe and Thoms, Johannes},
  journal={Statistics and computing},
  volume={18},
  number={4},
  pages={343--373},
  year={2008},
  publisher={Springer}
}

@article{roberts1998optimal,
  title={Optimal scaling of discrete approximations to Langevin diffusions},
  author={Roberts, Gareth O and Rosenthal, Jeffrey S},
  journal={Journal of the Royal Statistical Society: Series B (Statistical Methodology)},
  volume={60},
  number={1},
  pages={255--268},
  year={1998},
  publisher={Wiley Online Library}
}

@article{gorman1988analysis,
  title={Analysis of hidden units in a layered network trained to classify sonar targets},
  author={Gorman, R Paul and Sejnowski, Terrence J},
  journal={Neural networks},
  volume={1},
  number={1},
  pages={75--89},
  year={1988},
  publisher={Elsevier}
}

@article{pagani2022n,
  title={An n-dimensional Rosenbrock distribution for Markov chain Monte Carlo testing},
  author={Pagani, Filippo and Wiegand, Martin and Nadarajah, Saralees},
  journal={Scandinavian Journal of Statistics},
  volume={49},
  number={2},
  pages={657--680},
  year={2022},
  publisher={Wiley Online Library}
}

@article{ramsay2007parameter,
  title={Parameter estimation for differential equations: a generalized smoothing approach},
  author={Ramsay, Jim O and Hooker, Giles and Campbell, David and Cao, Jiguo},
  journal={Journal of the Royal Statistical Society Series B: Statistical Methodology},
  volume={69},
  number={5},
  pages={741--796},
  year={2007},
  publisher={Oxford University Press}
}

\end{document}